\documentclass[aps,prl,floatfix,showpacs,twocolumn]{revtex4}
\usepackage{epsfig}
\begin{document}

%\twocolumn[
%\hsize\textwidth\columnwidth\hsize\csname@twocolumnfalse\endcsname

\title{Theory of magnetic bipolar transistors}

\author{Jaroslav Fabian$^1$, Igor \v{Z}uti\'{c}$^2$, and S. Das Sarma$^2$ }
\affiliation{Institute for Theoretical Physics, Karl-Franzens University, 
Universit\"atsplatz 5, 8010 Graz, Austria \\
Condensed Matter Theory Center, Department of Physics, University of Maryland at  College
Park, College Park, Maryland 20742-4111, USA}

%\maketitle

\begin{abstract}
The concept of a magnetic bipolar transistor (MBT) is introduced.
The transistor has at least one magnetic region (emitter, base, or collector) characterized
by spin-splitting of the carrier bands. In addition, nonequilibrium (source) spin in MBTs can be  
induced by external means (electrically or optically). The theory of 
ideal MBTs is developed and discussed in the forward active regime where the 
transistors can amplify signals. It is shown that source spin can be injected from 
the emitter to the collector.  It is predicted that electrical current gain (amplification) can be
controlled effectively by magnetic field and source spin.
\end{abstract}
%]
%\newpage
\maketitle

Ideally, 
novel electronics applications build on the existing technologies with as
little added complexity as possible, while providing greater capabilities and
functionalities than the existing devices. Such is the promise of semiconductor
spintronics~\cite{DasSarma2001:SSC}
which aims at developing novel devices--utilizing electron spin, 
in addition to charge--which would provide spin and magnetic control of electronics 
and, {\it vice versa},  electronic control over spin and magnetism. Potential applications of 
semiconductor spintronics range from nonvolatile computer memories 
to spin-based quantum computing~\cite{DasSarma2001:SSC}. 
One particular promising implementation of semiconductor
spintronics is bipolar spintronics~\cite{DasSarma2001:SSC} which combines spin and charge
transport of both
electrons and holes in (generally magnetic) semiconductor heterostructures to control 
electronics.  In this Letter we propose a novel 
device scheme--magnetic bipolar (junction) transistor (MBT)--which, while in design a minor 
modification of the existing charge-based hetorojunction transistor (in fact, materials 
needed to fabricate MBTs are already available), has a great potential for extending 
functionalities of the existing device structures, since, as is demonstrated
here, its current gain (amplification) characteristics can be controlled by 
magnetic field and spin.

As semiconductor spintronics itself, bipolar spintronics still relies rather on 
experimentally demonstrated fundamental physics concepts 
(such as spin injection~\cite{Hammar1999:PRL,Fiederling1999:N,Ohno1999:N,Jonker2000:PRB}, 
spin filtering~\cite{Hao1990:PRB}, or semiconductor ferromagnetism~\cite{Ohno1998:S}) 
than on demonstrated working devices. But
the recent experiments ~\cite{Kohda2001:JJAP,Johnston-Halperin2002:PRB} on spin injection
through bipolar tunnel junctions clearly prove the potential of spin-polarized bipolar 
transport for both interesting fundamental physics and useful technological applications. 
We have recently shown theoretically that indeed spin-polarized  bipolar transport 
is a source of novel physical effects and
device concepts~\cite{Zutic2001:PRB,Zutic2001:APL,Zutic2002:PRL,
Fabian2002a:PRB}. In particular, we have analyzed the properties  of magnetic junction diodes,
demonstrating spin injection, spin capacitance, giant magnetoresistance, and a 
spin-voltaic effect. Here we formulate an analytic approach  
to study magnetic bipolar transistor (which is a very different structure from
the earlier spin transistors ~\cite{Datta1990:APL,Monsma1995:PRL}),
incorporating two magnetic {\it p-n} junctions in sequence. 
The step from a diode to a transistor is nontrivial conceptually
as it introduces new phenomena, most notably current amplification.  
Our two major findings are: source spin can be injected across a transistor
and electrical gain can be controlled by spin and magnetic field.

\begin{figure}
\centerline{\psfig{file=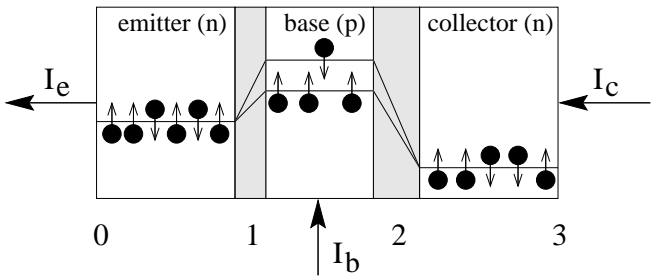,width=1\linewidth}}
\caption{Scheme of a magnetic $npn$ transistor. Shown are
conduction bands in each region. Only the base is magnetic
in the figure. The forward bias applied to junction 1
lowers the electrostatic barrier for electrons to cross from the
emitter to the base, while the reverse bias in junction 2 
increases the barrier in that junction. The shaded regions
around the junctions are the depletion layers. 
}
\label{fig:1}
\end{figure}

A scheme of MBT is shown in Fig.~\ref{fig:1}. We consider an $npn$ transistor with
spin-split conduction bands (the splitting is proportional to magnetic field and
is amplified by magnetic doping) and with source spin (which is incorporated here through
boundary conditions) injected, in principle, to any region. Source spin, in addition
to applied bias, brings about nonequilibrium carrier population and thus electrical
current.  In the following we generalize the theory developed for magnetic 
{\it p-n} junctions~\cite{Fabian2002a:PRB} 
to study magnetic transistor structures.
All the assumptions of that theory apply here. Most important, carriers obey 
nondegenerate Boltzmann statistics,  
nonequilibrium carrier densities are smaller than the doping densities (the low injection
or low bias limit), and carrier recombination and spin relaxation is neglected in the 
depletion layers. Further, we express voltages in the units of thermal energy 
$k_BT$, and make them positive for forward biasing. 

Our first task is to obtain the electron and spin densities at the two depletion layers.
Once these are known, the density profiles can be calculated
using the formulas provided in Tab. II of Ref.~\onlinecite{Fabian2002a:PRB}. In the following
the quantities at the emitter-base (collector-base) depletion layer edges carry index
1 (2). To  simplify 
complex notation we adopt terminology that is useful in treating an 
arbitrary array of magnetic {\it p-n} junctions, though here we limit ourselves 
to MBT which is the smallest nontrivial array
of such kind. We denote by scalar $u$ the nonequilibrium spin 
density in the $n$ regions (here emitter $e$ and collector $c$), 
and by vector $\bf v$ the nonequilibrium
electron (the first component) and spin (the second component) densities in 
the $p$ regions (here only base $b$). The boundary conditions are specified
by $u$ and $\bf v$ at the emitter and collector contacts to the external electrodes. 
In our case the boundary spin densities are $u_0$ and $u_3$ which are to be treated 
as input parameters. The notation inside the
array follows the indexing of the junctions. For example, ${\bf v}_2$ is the
nonequilibrium density vector in the $p$ side at the second depletion layer edge
(in our case it is the density in the base at the $b-c$ depletion layer). The
values of $u_1$, ${\bf v}_1$, etc. need to be obtained self-consistently 
requiring \cite{Fabian2002a:PRB}
that the (spin-resolved) chemical potentials
and spin currents are continuous across the depletion layers.
The following is the basic set of equations describing the coupling of charge and spin
(the coupling is both intra- and inter-junction) in
the magnetic transistor system  ~\cite{Fabian2002a:PRB}:  
\begin{eqnarray} \label{eq:u1}
u_1&=&\gamma_{0,1} u_0 +{\bf C}_1\cdot {\bf v}_2, \\ \label{eq:v1}
{\bf v}_1& =& {\bf v}_1^0 + {\bf D}_1 u_1,
\end{eqnarray}
for junction $1$, and
\begin{eqnarray} \label{eq:u2}
u_2&=&\gamma_{0,2} u_3 +{\bf C}_2\cdot {\bf v}_1, \\ \label{eq:v2}
{\bf v}_2& =& {\bf v}_2^0 + {\bf D}_2 u_2,
\end{eqnarray}
for junction 2. The notation goes as follows. For a general junction
${\bf v}^0=[\exp(V)-1](n_{0p},s_{0p})$ is the nonequilibrium
density vector due to applied bias (across the junction) $V$ (but no source spin), 
${\bf C}=[\alpha_{0p}(\gamma_2-\gamma_1),\gamma_1]$, and 
\begin{equation}
{\bf D}=\frac{n_{0p} e^V}{N_d}\frac{1}{1-\alpha^2_{0n}}
\left (\alpha_{0p}-\alpha_{0n},1-\alpha_{0p}\alpha_{0n}\right ).
\end{equation}
Symbol $n_{0p}$ ($s_{0p}$) stands for the electron (spin) equilibrium density in the 
$p$ region of the junction, $N_d$ is the donor doping density of the $n$-region,
and $\alpha_{0n}$ ($\alpha_{0p}$) is the equilibrium electron spin polarization (the
ratio of spin and electron density) in the $n$ ($p$) region adjacent to the junction. 
The geometric/transport factors $\gamma_0$ through $\gamma_2$ 
are determined from carrier diffusivities, carrier recombination and spin relaxation times, and
effective widths of the adjacent bulk regions~\cite{Fabian2002a:PRB}. 
 We note that equations analogous to 
Eqs.~\ref{eq:u1}--\ref{eq:v2} can be written for holes, if their polarization is taken
into account. The solution to Eqs.~\ref{eq:u1}--\ref{eq:v2} is
\begin{equation} \label{eq:solution}
u_2=\gamma_{0,1}\left ({\bf C}_2\cdot{\bf D}_1 \right )u_0 
+ \gamma_{0,2} u_3 + {\bf C}_2\cdot {\bf v}^0_1,
\end{equation}
where we have neglected terms of order $[n_{0p}\exp(V)/N_d]^2$, consistent
with the small
injection limit. The formulas for $u_1$, ${\bf v}_1$, and ${\bf v}_2$ can be
obtained directly by substituting Eq.~\ref{eq:solution} for $u_2$ into 
Eqs.~\ref{eq:u1} through \ref{eq:v2}.

Equation \ref{eq:solution} describes spin injection through MBT, since $u_2$ is
the nonequilibrium spin in the collector at the depletion layer with the base. The
first term on the right-hand side (RHS) of Eq.~\ref{eq:solution} 
represents transfer of source spin $u_0$
from the emitter to the collector. Indeed, for a nonmagnetic transistor (the equilibrium
spin polarizations are zero) the
transferred source spin is $u_3=\gamma_{0,1}\gamma_{1,2}n_{0b}\exp(V_1)u_0$. Here
$\gamma_0$ describes the transfer of source spin through the emitter--majority
carrier spin injection. Once the spin is in the base, it becomes the spin of the 
minority carriers [hence the minority density factor $n_{0b}\exp(V_1)$],
diffusing towards the $b-c$ depletion layer. The built-in electric field in 
this layer sweeps the spin into the collector, where it becomes the spin of
the majority carriers again, by the process of minority-carrier spin 
pumping~\cite{Zutic2001:PRB,Fabian2002a:PRB}. 
Can the
injected spin polarization in the collector be greater than the source spin polarization? 
The answer is negative in the low-injection regime. It would be tempting to let the
spin diffusion length in the collector to increase to large values to get 
a greater pumped spin. But that would increase the importance of electric
field in the $n$-regions and the theory (which is based on charge and spin diffusion and
not spin drift) would cease to be valid. However, the spin density in the collector
can be greater than that in the base (as illustrated in the example below), 
demonstrating that spin spatial decay is not, in general, monotonically
decreasing. The second term on the RHS of Eq.~\ref{eq:solution} results from diffusion
of the source spin in the collector (described by $\gamma_{0,2}$). 
Finally, the third term, which is independent of source spin, 
results from the (intrinsic) spin pumping by the minority
channel of {\it nonequilibrium} spin generated in the base by the forward current
through junction 1. This term vanishes if the base is nonmagnetic ($\alpha_{0b}=0$).

\begin{figure}
\centerline{\psfig{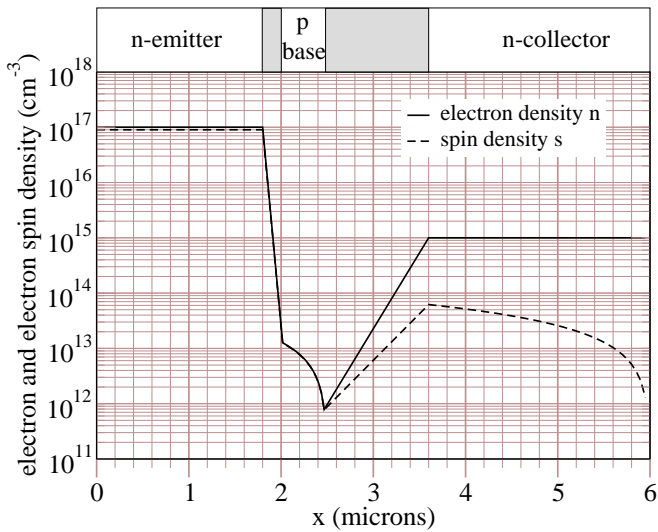}}
\caption{Calculated electron and spin density profiles in a Si-based
$npn$ transistor with magnetic base and source spin in the emitter. 
The transistor geometry is shown at the top. The densities inside
the depletion layers are not calculated, and are shown here (with
no justification beside guiding the eye) as straight lines connecting 
the densities at the depletion layer edges.
}
\label{fig:2}
\end{figure}

To illustrate spin injection across MBT we plot in Fig.~\ref{fig:2} the calculated electron 
and spin density
profiles in a Si-based magnetic $npn$ transistor with magnetic base (and nonmagnetic
emitter and collector) and with source spin in the emitter. The geometry of the device is 
depicted at the top of the figure.
The emitter, base, and collector are doped (respectively) with $N_{e}=10^{17}$, 
$N_{b}=10^{16}$, and $N_c=10^{15}$ donors, acceptors, and donors per cm$^3$. The carrier
and spin relaxation times are taken to be 0.1 $\mu$s (it is not clear what spin relaxation
times of conduction electrons in Si should be~\cite{Jia1996:IEEE},
but due to the small spin-orbit coupling
they are expected to be on the order of sub-microseconds, 
rather than sub-nanoseconds
as in GaAs), the electron (hole) diffusivities
are $D_n=100$ ($D_p=10$) cm$^2$s$^{-1}$, the dielectric constant is 12, and the intrinsic 
carrier concentration is $n_i=10^{10}$ cm$^{-3}$. The transistor is at
room temperature. The applied biases are $V_1=V_{be}=0.5$ volts and $V_2=V_{bc}=-0.2$ volts.
The spin splitting of the base conduction band is $2\zeta_b=2$ 
(in $k_B T$), yielding
the equilibrium spin polarization $\alpha_{0b}=\tanh(\zeta_b)=0.76$; the
source spin polarization at the emitter (at $x=0$) is $u_0/N_D=0.9$.
Finally, we assume charge and spin ohmic contact at $x=3$, meaning that both
carrier and spin densities are at their equilibrium levels.
Figure \ref{fig:2} demonstrates that spin injection is possible all the way 
from the emitter, through the base, down to the collector. The density of
the injected spin in the collector depends on many factors, most notably
on the forward bias $V_{be}$ and on the spin diffusion lengths in the base and in the collector. 
The spin density (but not spin polarization) even increases as one goes from 
the base to the collector, consistent with our notion of spin 
amplification~\cite{Zutic2001:PRB,Fabian2002a:PRB}. 
The injected spin 
polarization $u_2/N_c$ in the above example is about 2\%, but it would be greater for higher
$V_{be}$, longer spin relaxation times, and smaller base widths.

We now turn to the question of current gain (amplification) and its control
by magnetic field (through $\zeta_b$) and source spin (through $\delta \alpha_e$, the
nonequilibrium spin polarization in the emitter at junction 1).  
Electric currents are readily evaluated once the nonequilibrium
carrier densities at the depletion layers are known. Thus the emitter
current 
\begin{equation}
j_{e}=j^n_{gb}\frac{\delta n_{be}}{n_{0b}}-j^n_{gb}\frac{1}{\cosh(\tilde{w}_b/L_{nb})}
\frac{\delta n_{bc}}{n_{0b}}+ j^p_{ge}\frac{\delta p_{be}}{p_{0e}},
\end{equation}
and the collector current
\begin{equation}
j_{c}=-j^n_{gb}\frac{\delta n_{bc}}{n_{0b}}+j^n_{gb}\frac{1}{\cosh(\tilde{w}_b/L_{nb})}
\frac{\delta n_{be}}{n_{0b}}+j^p_{gc}\frac{\delta p_{bc}}{p_{0c}},
\end{equation}
where we denote the generation currents for electrons and holes (with
the indexing of the appropriate region) respectively as~\cite{Fabian2002a:PRB}
$j^n_g=(qD_{n}/L_n)n_0 \coth (\tilde{w}/L_n)$, and 
$j^p_g=(qD_{p}/L_p)p_0 \coth (\tilde{w}/L_p)$. 
Here $q$ is the proton charge, $L_n$ ($L_p$) is the electron (hole) minority
diffusion length, and $\tilde{w}$ is the effective (taking into account bias
variation of the depletion layer widths) width of the
region; $\delta n$ ($\delta p$) are the nonequilibrium electron (hole) 
densities at the corresponding depletion layer. 
In the active control regime ($V_{be}>0$ and $V_{bc} < 0$) the hole collector current and the current
driven by the nonequilibrium density $\delta n_{bc}$ becomes negligible.
Finally, the base current is given by the current continuity (see Fig. \ref{fig:1})
as $j_b=j_e-j_c$.

The current amplification factor $\beta$ is the ratio of the collector current 
to the base current (if $\beta$ is large, typically about 100, small changes
in $j_b$ lead to large variations in $j_c$). For illustration we consider only
the case of magnetic base and emitter source spin, and consider (as is typically
done in transistor physics) thin bases ($\tilde{w}_b \ll L_{nb}, L_{sb}$ where $L_{sb}$ 
is the spin diffusion length in the base). The gain of MBT can then be written 
as $\beta=1/(\alpha_T' + \gamma')$, where
\begin{equation}
\alpha_T'= (\tilde{w}_b/L_{nb})^2/2,
\end{equation}
and
\begin{equation}
\gamma'=\frac{N_bD_{pe}}{N_eD_{nb}}\frac{\tilde{w}_b}{L_{pe}\tanh(\tilde{w}_e/L_{pe})}
\frac{1}{\cosh(\zeta_b)(1+\delta \alpha_e \alpha_{0b})}.
\end{equation}
The two factors $\alpha_T'$ and $\gamma'$ are related to the usual base transport $\alpha_T$
and emitter efficiency factor $\gamma$ by $\alpha_T=1/(1+\alpha_T')$ and $\gamma=1/(1+\gamma')$.
They represent, respectively,  the contribution to the gain by the carrier recombination
in the base and by the efficiency of the electrons injected by the emitter to carry the total
charge current in the emitter (for a standard reference on nonmagnetic transistors 
see, for example, ~\cite{Tiwari:1992}). In MBT the base transport cannot be controlled
by either spin or magnetic field, since it is related only to carrier recombination
in the base (one can, however, consider more specific cases where $L_{nb}$ depends
on $\zeta_b$, in which case even $\alpha_T'$ could be controlled). The emitter
efficiency, on the other hand, varies strongly with both $\zeta_b$ and 
$\delta \alpha_{be}$.

Under what circumstances can we control $\beta$ by magnetic field and spin most
effectively? The answer lies in the relative magnitudes of $\alpha_T'$ and
$\gamma'$. In GaAs-base transistors the two might have similar amplitudes, 
since the carrier recombination is rather fast, although additional band
structure engineering (making heterojunctions) usually significantly
enhances $\gamma'$ at which point $\gamma'$ might dominate. The situation
is much more favorable in Si (or Si/Ge) based transistors, which have long
carrier recombination times and it is the emitter efficiency $\gamma'$ which
determines the gain. In this case $\beta=1/\gamma'$ and
\begin{equation} \label{eq:gain}
\beta \sim \cosh(\zeta_b)(1+\delta \alpha_{be}\alpha_{0b}).
\end{equation}
The gain varies exponentially with $\zeta_b$ and is asymmetrically modulated
by the magnetic field, depending on the relative orientation of the magnetic
field and source spin polarization. The physics behind Eq.~\ref{eq:gain}
is quite illuminating. The emitter efficiency is the ratio of the electron
emitter current to the total emitter current (which includes the hole current). The
electron part of the current depends linearly on the electron minority
carrier density in the base. This density is modulated, separately, by $\zeta_b$,
which changes the effective band gap in the material and thus
the equilibrium minority carrier density--according to $n_{0b}=n_i^2\cosh(\zeta_b)/N_b$
\cite{Zutic2002:PRL,Fabian2002a:PRB}, and by the amount of nonequilibrium spin 
(through the spin-voltaic effect~\cite{Zutic2002:PRL,Johnson1985:PRL}). 
Similar control of gain could be achieved by having a magnetic
emitter. In such a case it would be the equilibrium minority hole density (and thus
the hole emitter current) which would be modified by magnetic field, changing the
emitter efficiency. All the effects associated with the conduction band
spin splitting can be also observed when the splitting is (also) in the 
valence band.

\begin{figure}
\centerline{\psfig{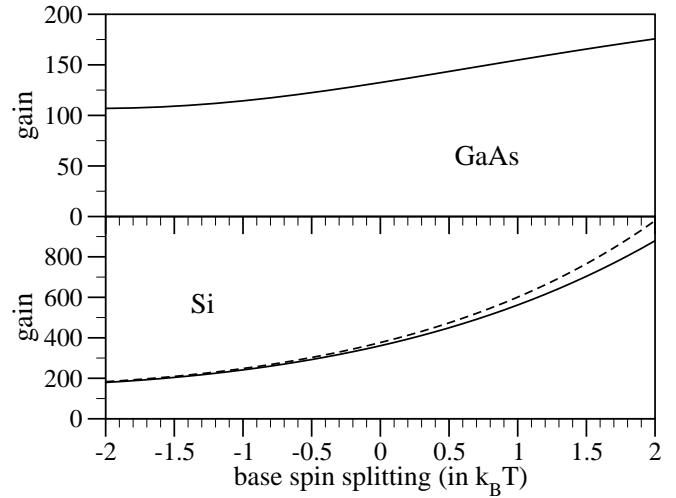}}
\caption{Calculated gain of a magnetic $npn$ transistor with magnetic base
and source spin in the emitter. The upper (lower) graph is for GaAs (Si)
materials parameters. The dashed line in the Si graph is the contribution
of the emitter efficiency which controls the current gain in Si-based
transistors.
}
\label{fig:3}
\end{figure}

To illustrate the gain control of magnetic field and spin we 
calculate $\beta$ for the same $npn$ geometry as in Fig.~\ref{fig:2}, but 
now with two different sets of materials parameters. Figure ~\ref{fig:2},  
top part,
is for GaAs (with $n_i=1.8\times 10^{6}$ cm$^{-3}$, 
dielectric constant of 11, and recombination and relaxation times of 1 ns,
keeping all the other parameters unchanged), while the bottom part is for Si.  
The calculated gain as a function of conduction band
spin splitting $\zeta_b$ in the base is shown in Fig.~\ref{fig:3}. The source
spin polarization $u_0/N_d$ at the emitter is set to 0.9 (which is roughly
also $\alpha_{be}$). The figure shows that current
gain (amplification) is significantly influenced by magnetic field (which
controls the splitting), but much more in Si than in GaAs, for the reasons stated
earlier. 

Magnetic bipolar transistor could be also called magnetic heterostructure
transistor. Indeed, MBT's functionality is based on tunability 
of electronic properties by band structure engineering. In contrast to the 
standard (nonmagnetic) heterostructure transistors, however, MBT's band structure (the spin-split
conduction band) is not a fixed property, but can change on demand, during the device operation, 
by changing the
magnetic field. One can also have magnetic heterostructure transistors
with variable spin splitting in the base producing magnetic drift \cite{Fabian2002a:PRB} of
the spin carrying minority carriers (as in drift-base
transistors) to further enhance spin current and the resulting spin injection 
into the collector. Interesting effects could be observed by using ferromagnetic
semiconductors for the base. Similarly to optical induction of ferromagnetism
by optical injection of carriers~\cite{Koshihara1997:PRL}, 
emitter can inject (presumably in the high injection limit which goes beyond
the scope of our theory) high density carriers into the base, changing the base's magnetic
state (on and off, depending on the density of the nonequilibrium minority
electrons, or twisting the magnetic moment orientation, if the injected
electrons are spin-polarized). 
This could be an alternative electronic way of switching (or modifying) 
semiconductor ferromagnetism~\cite{Ohno2000:N,Park2002:S}, which 
could lead to numerous novel functionalities. 

This work was supported by DARPA, the NSF-ECS, and the US ONR. 

\bibliographystyle{apsrev}
\bibliography{spintronics}
%\bibliography{refrmp}

\end{document}